\documentclass[a4paper, 11pt]{article}
\pdfoutput=1
\usepackage{graphicx}
\usepackage{epsf,amsmath,bbold,amsfonts,stmaryrd}

\usepackage[utf8]{inputenc}
\usepackage{mathrsfs}
\usepackage{appendix}
\usepackage{amssymb}
\usepackage{float}
\usepackage{color}
\usepackage{hyperref}
\hypersetup{pageanchor=false,hidelinks} 
\usepackage{indentfirst}
\usepackage{url}
\usepackage{xurl}
\usepackage{caption}
\usepackage[numbers,square,comma,sort&compress,merge]{natbib}

\usepackage{subfigure}

\numberwithin{equation}{section}
\let\originalleft\left
\let\originalright\right
\renewcommand{\left}{\mathopen{}\mathclose\bgroup\originalleft}
\renewcommand{\right}{\aftergroup\egroup\originalright}
\mathcode`\*="8000
{\catcode`\*=\active\gdef*{\mathclose{}\,\mathopen{}}}

\newcommand{\be}{\begin{equation}}
\newcommand{\ee}{\end{equation}}
\newcommand{\bea}{\setlength\arraycolsep{2pt} \begin{eqnarray}}
\newcommand{\eea}{\end{eqnarray}}
\newcommand{\nn}{\nonumber}

\def \nn {\nonumber}
\def\degree{${}^{\circ}$}

\newcommand{\rI}{r_\text{ISCO}}

\begin{document}
\title{Polarized images of orbiting charged particles in Schwarzschild--Melvin spacetime}

\author{
Yuheng Du, Hu Zhu, Sijie Gao$^{\ast}$, Minyong Guo$^{\ast}$}
\date{}

\maketitle

\vspace{-10mm}

\begin{center}
{\it
School of Physics and Astronomy, Beijing Normal University, Beijing 100875, China\\\vspace{2mm}
Key Laboratory of Multiscale Spin Physics (Beijing Normal University), Ministry of Education, Beijing 100875, China
}
\end{center}

\vspace{8mm}

\begin{abstract}
We revisit the innermost stable circular orbits (ISCOs) of charged particles and study the polarized images of synchrotron radiations emitted from such orbiting hotspots on the equatorial plane in {Schwarzschild}-Melvin black hole spacetime. We obtain a constraint on the magnetic field {for the existence of charged-particle circular orbits}. In particular, we identify a critical value for the strength of the magnetic field $B_c$ that the circular orbits can have positive and negative angular momentums {below} $B_c$ while only {the negative angular momentum branch survives above} $B_c$. Furthermore, we investigate and discuss the primary and secondary images of circularly orbiting charged hotspots carrying the information of polarization directions observed by distant observers. {We also find that the observed polarization angle varies rapidly with source azimuth when the emitted photon direction is nearly aligned with the local magnetic field, and explain the origin of this behavior.}
\end{abstract}

\vfill{\footnotesize $~~~~~~\ast$ Corresponding author: sijie@bnu.edu.cn, minyongguo@bnu.edu.cn}

\newpage
\baselineskip 18pt
\section{Introduction}\label{sec1}
The Event Horizon Telescope (EHT) Collaboration have recently released a polarized image of photon rings around the supermassive black hole at the centre of M87 galaxy \cite{EventHorizonTelescope:2021bee, EventHorizonTelescope:2021srq}, which reveals that a significant magnetic field exists in the vicinity of M$87^\ast$. Surrounding charged particles would be affected by the magnetic field and strong gravitational field simultaneously so that lots of interesting phenomena emerge in the environment containing a black hole. Moreover, the reason why EHT is able to observe polarized photon ring structures is that electrons accelerated by a magnetic field emit synchrotron radiations at high velocities and synchrotron radiations exhibit excellent polarization properties. The converse also applies: from the polarization information encoded in the polarized image of photon rings we can infer the magnetic field configuration for a given spacetime.

In order to understand the observations theoretically, simulations of polarized emission has played an essential role in the study of astrophysics \cite{Bromley:2001er, Broderick:2003fc, Li:2008zr, 10x,101526, Moscibrodzka:2017gdx, Gold:2016hld, Marin:2017rlw, Jimenez-Rosales:2018mpc, Huang:2024bar, Palumbo_2020, Moscibrodzka:2019adb,Feng:2026gob,Zeng:2026gly,Hu:2026vjh}. But beyond that, some analytical models are also presented to gain an insight into the polarized black hole image. Along this line, an analytical treatment of high-spin Kerr black hole spacetime was proposed in \cite{Lupsasca:2018tpp}. In \cite{EventHorizonTelescope:2021btj}, EHT Collaboration developed a model of an equatorial magnetized fluid orbiting a Schwarzschild black to understand polarimetric images of black holes. Soon, this model has been generalized to a Kerr black hole spacetime and other black hole geometries arise from modified theory of gravity in \cite{Gelles:2021kti, Qin:2021xvx, Zhang:2021hit}. Also, In virtue of such a model, one can investigate the polarized image for arbitrary emission radius, magnetic field geometry, equatorial fluid velocity, and observer inclination to match the astroobservation. However, such models are not particularly satisfactory for physicists, since some parameters in such models are empirical rather than determined by the basic physical equations. Recently, in \cite{Hu:2022sej}, authors built a toy model to study the  polarized images of synchrotron radiations of charged hotspots (or thin free electron gas, equivalently) in Schwarzschild black hole spacetime. This model is obviously very simplified by ignoring the thermal motions and interactions of fluids, and thus, the calculated results are not comparable with real observational data. Nevertheless, each detail of the model is based on fundamental physical laws and no parameters are manually placed. As a result, this model can be seen as a first step and must be an important element of the final complete theory. In addition, this model offers an efficient way to  decouple and characterize the effects of the magnetic and spacetime on black hole images over a complex case of multiple parameter coupling.

More precisely, in \cite{Hu:2022sej} they focused on the Wald magnetic fields, that is, the Schwarzschild black hole is immersed in vertical magnetic fields, which satisfy the Maxwell equation and the solution is firstly found by Wald in \cite{Wald:1974np}. Moreover, the Wald magnetic field are assumed to have no back reaction to the spacetime geometry. On the other hand, if a vertical magnetic field becomes strong and the back reaction cannot be ignored, {one exact model is the} Schwarzschild-Melvin black hole (SMBH) which is an exact solution of Einstein-Maxwell theory found by Ernst \cite{Ernst:1976}. The SMBH spacetime is not asymptotically flat due to the existence of magnetic field at infinity and the SMBH can be regarded as a Schwarzschild black hole embedded in a Melvin universe \cite{Melvin:1963qx}, so that, this type of magnetic fields is also dubbed Melvin magnetic fields. The SMBH has been studied in various aspects, such as the motion of charged particles, neutral particles and photons on the equatorial plane \cite{Dadhich_1979, 1992Chaotic, 2010Motion, Lim:2015oha, 2007Geodesics, 1983Null, Stuchlik:1999mro}, shadows and lensing of SMBH \cite{Junior:2021dyw, Wang:2021ara}. In this work, we would like to investigate the influence of the Melvin magnetic fields on the polarized image of synchrotron radiations emitted from an equatorial charged hotspot. The source motion is described by the charged circular orbits. The emitted polarization vector is determined by the local magnetic field and photon momentum in the emitter's comoving frame, and is then propagated to the observer by numerical parallel transport along null geodesics. We assume the charged particles to move in a circular orbit {at and outside} the innermost stable circular orbit (ISCO) on the equatorial plane.
Thus, our research starts with the ISCO in the SMBH spacetime and we find some interesting properties of ISCOs that were not found in the literature passingly. Nevertheless, our main attention is focused on the shape of the image and the polarization directions of the synchrotron radiations of the orbiting source.

The remaining part of this paper is organized as follows. In Sec. \ref{sec2}, we introduce the SMBH spacetime and revisit the ISCO of charged particles. In Sec. \ref{sec3}, we set up the methods for calculating the polarized image of synchrotron radiations emitted from charged particles of circular motion observed by distant observers in SMBH spacetime. In Sec. \ref{sec4}, we show the details of results. In Sec. \ref{sec5}, we summarize and conclude this work.

\section{SMBH spacetime and ISCO of charged particles}\label{sec2}
In this section, we would like to give a necessary review of the SMBH spacetime and revisit the ISCO of charged particles in this background. In the {Schwarzschild-like coordinates} $\{t, r, \theta, \phi\}$, the metric of the SMBH spacetime reads\cite{Ernst:1976}
\begin{equation}\label{metric}
d s^{2}=-\Lambda^{2}(r, \theta)f(r) d t^{2}+\frac{\Lambda^{2}(r, \theta)}{f(r)} d r^{2} +r^{2} \Lambda^{2}(r, \theta) d \theta^{2}+\frac{r^{2} \sin ^{2} \theta}{\Lambda^{2}(r, \theta)} d \phi^{2}\,,
\end{equation}
where
\begin{equation}\label{2}
\Lambda(r, \theta)=1+\frac{B^{2} r^{2} \sin ^{2} \theta}{4}\,, \quad\quad f(r)=1-\frac{2 M}{r}\,,
\end{equation}
$M$ is the mass of the BH and $B$ is {the background magnetic-field parameter}. The corresponding electromagnetic potential takes in this form
\begin{equation}\label{mpo}
A_a=\frac{B r^{2} \sin ^{2} \theta}{2 \Lambda} (d \phi)_a\,.
\end{equation}
It's worth noting that {the stress-energy of the magnetic field is included in the SMBH solution}, so that it has a non-trivial back reaction to the background spacetime. Moreover, one can notice that the SMBH is not asymptotically flat due to {the Melvin magnetic background}. When $B=0$, the SMBH spacetime reduce to the usual Schwarzschild spacetime and it returns to the so-called Melvin universe with $M=0$.  As long as $M$ is not equal to zero, Inevitably one always finds the coordinate $r=0$ to be the singularity of the spacetime and $r=2M$ is a coordinate singularity which corresponds to an apparent horizon and it is believed that there is a BH inside the apparent horizon. In the following, we would like to focus on the spacetime containing a SMBH that is, $M\neq0$. For the sake of convenience, we are going to set $M=1$ without loss of generality.

\begin{figure}[!htbp]
  \centering
  \includegraphics[width=8.0cm]{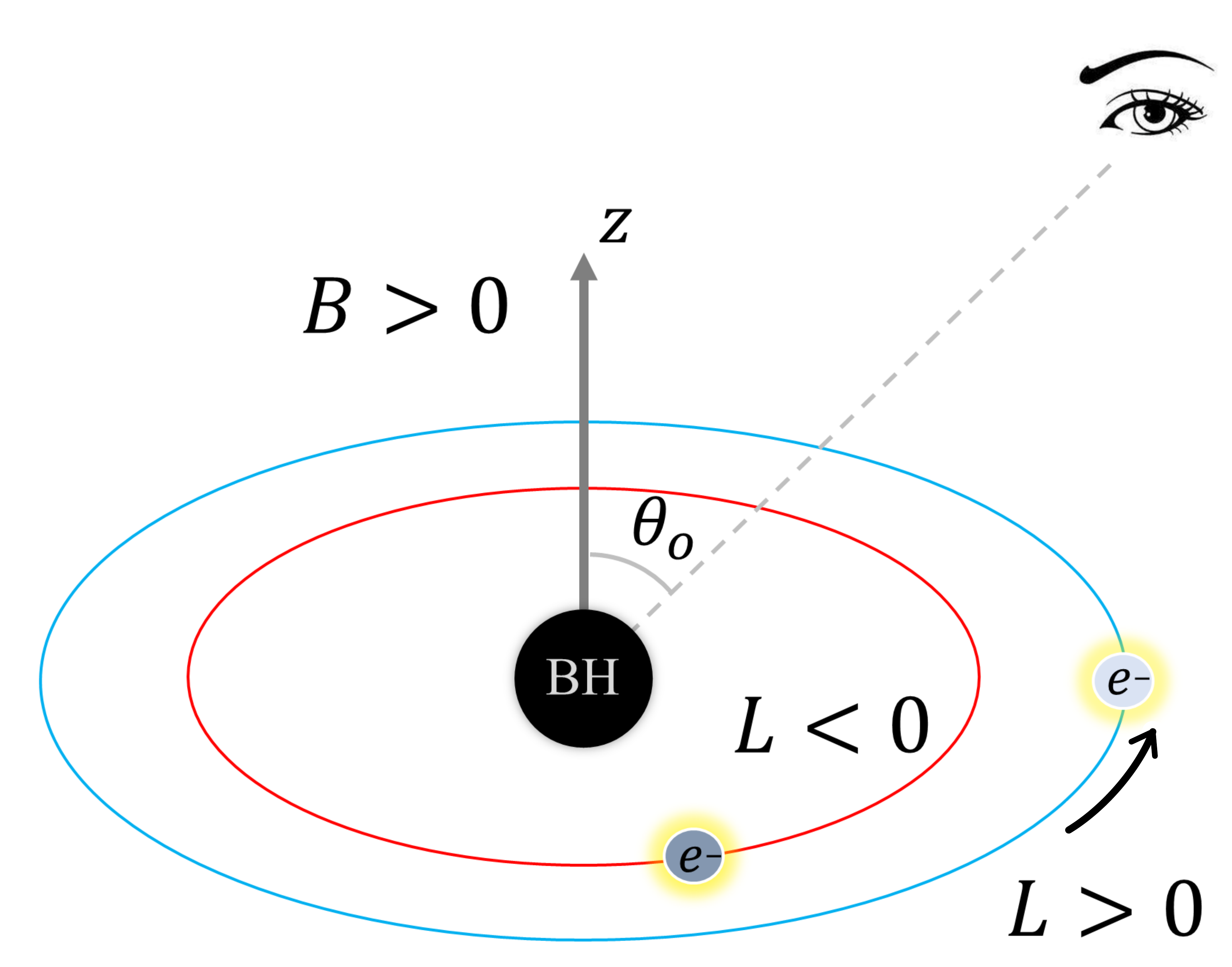}
  \caption{{}A diagram of the orbiting charged particles around the SMBH black hole on the equatorial plane, which includes borrowings in \cite{Hu:2022sej}. In our convention, we set $B>0$ and $e<0$.}\label{diam}
\end{figure}

Next, Let us consider the ISCO of charged particles moving in the SMBH spacetime. We assume the $4-$velocity of a charged particle is denoted by $Z^a$, which is in general given by
\bea
Z^a=\dot{t}\left(\frac{\partial}{\partial t}\right)^a+\dot{r}\left(\frac{\partial}{\partial r}\right)^a+\dot{\theta}\left(\frac{\partial}{\partial \theta}\right)^a+\dot{\phi}\left(\frac{\partial}{\partial \phi}\right)^a\,,
\eea
where overdots mean derivative with respect to the proper time $\tau$. Noting that $\left(\frac{\partial}{\partial t}\right)^a$ and $\left(\frac{\partial}{\partial \phi}\right)^a$ are Killing vectors of the SMBH spacetime, we have two corresponding conserved quantities,
\bea\label{tel}
E=-Z_t=\Lambda^{2}f\dot{t}\,,\quad\quad L=Z_\phi+eA_\phi=\frac{r^{2} \sin ^{2} \theta}{\Lambda^{2}} \dot{\phi}+e B \frac{r^{2} \sin ^{2} \theta}{2 \Lambda}\,,
\eea
which are the energy per unit mass and the angular momentum per unit mass of the charged particles, respectively. And, $e$ is the charge per unit mass of the particle. We want to stress that since charged particles would be acted upon by Lorentz force in the magnetic field, the term $eA_\phi$ appeared in Eq. (\ref{tel}) cannot be omitted, while this term needs to be dropped for neutral particles and photons \footnote{{In classical physics, photons experience no Lorentz force from magnetic fields;} however, if the effect of quantum electrodynamics is included, the dispersion relationship of photons changes and photons no longer travel along geodesics, see examples in \cite{Hu:2020usx, Zhong:2021mty}.}. On the other hand, due to the $\mathcal{Z}_2$ symmetry of the spacetime, one is allowed to choose $B>0$ to study the effects of magnetic fields on charged particles. Moreover, considering the invariance of Eqs. (\ref{metric}) and (\ref{mpo}) under the transformation $B\to-B\,,\phi\to-\phi$, in terms of studying the orbits of charged particles in a certain magnetic field, there is no essential difference between $e>0$ and $e<0$. Note that we are going to investigate the polarized image of synchrotron radiations of electrons, we might as well set $e<0$ in the following. A diagram is given in Fig. \ref{diam} to make the explanation to be clearer. In particular, we want to stress that for $L>0$ we have {$d\phi/dt>0$} from Eq. (\ref{tel}), while for $L<0$ the sign of $d\phi/dt$ is {uncertain}, since $d\phi/dt$ has the same sign with {$\left(L-\frac{eBr^2\sin^2\theta}{2\Lambda}\right)$} with $\frac{eBr^2\sin^2\theta}{2\Lambda}$ being negative in our convention. Thus, we would like to use the terminology of $L>0$ and $L<0$ orbits instead of retrograde and prograde orbits in this work.

In addition, in this work, we would like to focus our attentions on the equatorial orbits, thus, we have $\theta=\pi/2$. Then, from the normalization condition of the $4-$velocity, that is, $Z^aZ_a=-1$, we find
\bea
\Lambda^{4} \dot{r}^{2}=E^{2}-V(r)\,,
\eea
where
\bea\label{effp}
V(r) =\Lambda^{4} f\left[\frac{1}{r^{2}}\left(L-\frac{eB r^{2}}{2 \Lambda}\right)^{2}+\frac{1}{\Lambda^{2}}\right]\,,
\eea
is the so-called effective potential of the charged particles in the radial motion. Since $\dot{r}^2\ge0$ is always true, we always have $E^2-V(r)\ge0$ and the condition for equality gives the turning points in the radial motion. For circular orbits with fixed orbit radii, we have $r=\text{const}$ and $\dot{r}=0$, which correspond to $E^2-V(r)=0$ and $\partial_rV(r)=0$. Note that not all the circular orbits are stable, in order to find a stable {or marginally stable} circular orbit, the condition $\partial_r^2V(r)\ge0$ has to be imposed. Combining the equations $E^2-V(r)=0$, $\partial_rV=0$ and $\partial_r^2V=0$, one can obtain the radius of the ISCO. In order to know more details of the ISCO, we would like to study the behavior of the effective potential carefully.

\begin{figure}[!htbp]
  \centering
  \includegraphics[width=10.0cm]{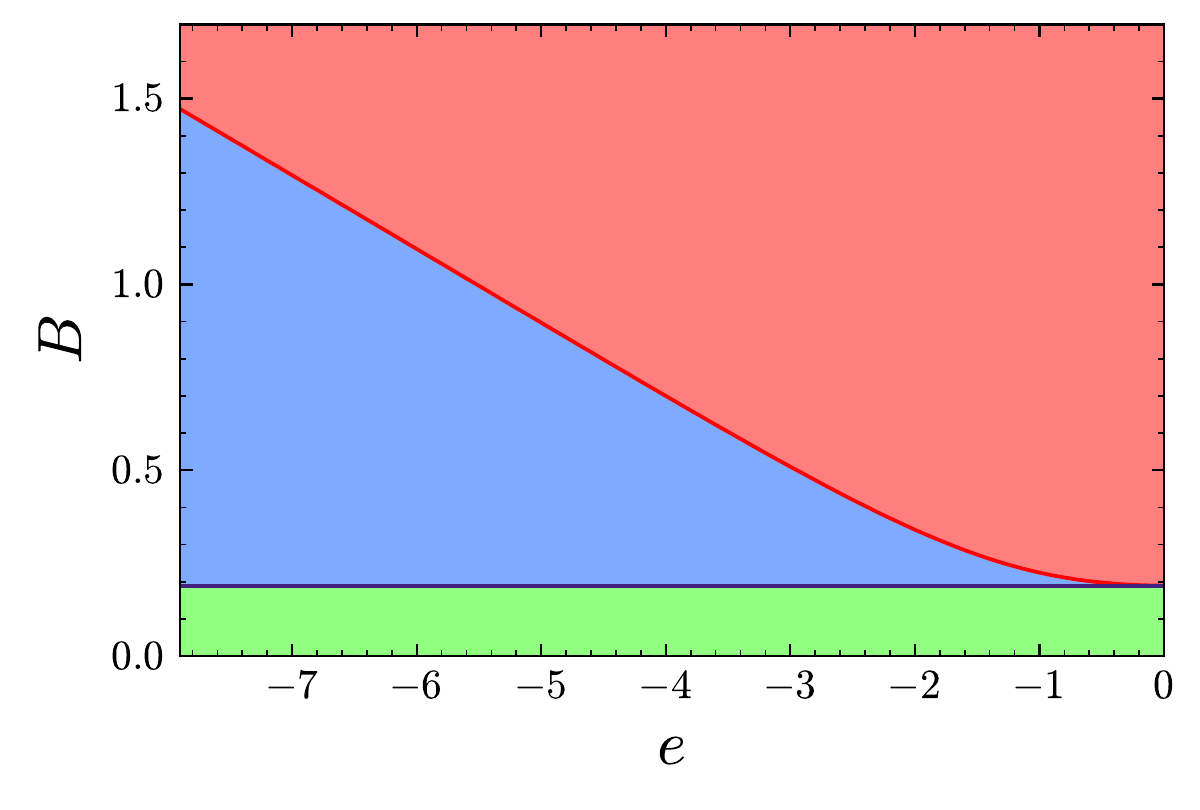}
  \caption{Constraints on $e$ and $B$ from the existence of physical solutions to the equation $\partial_rV=0$.}\label{constraint}
\end{figure}

{For a circular orbit, the energy is determined by $E^2=V(r)$. We therefore turn to the radial balance condition $\partial_rV=0$.} The explicit expression of $\partial_rV$ can be found in this form
\bea\label{pvl}
{\frac{32r^4\partial_r V}{\Lambda}}=XL^2(4+B^2r^2)^2-4Y e B Lr^2+4Zr^2
\eea
where
\bea
X&=&3B^2r^3-5B^2r^2-4r+12\,,\nn\\
Y&=&B^4r^4(3r-5)+4B^2r^2(3r-4)+16\,,\nn\\
Z&=&B^4e^2r^4(3r-5)+4B^2r^2[2r-3+(r-1)e^2]+16\,.
\eea

\begin{figure}[!htbp]
  \centering
  \includegraphics[width=10.0cm]{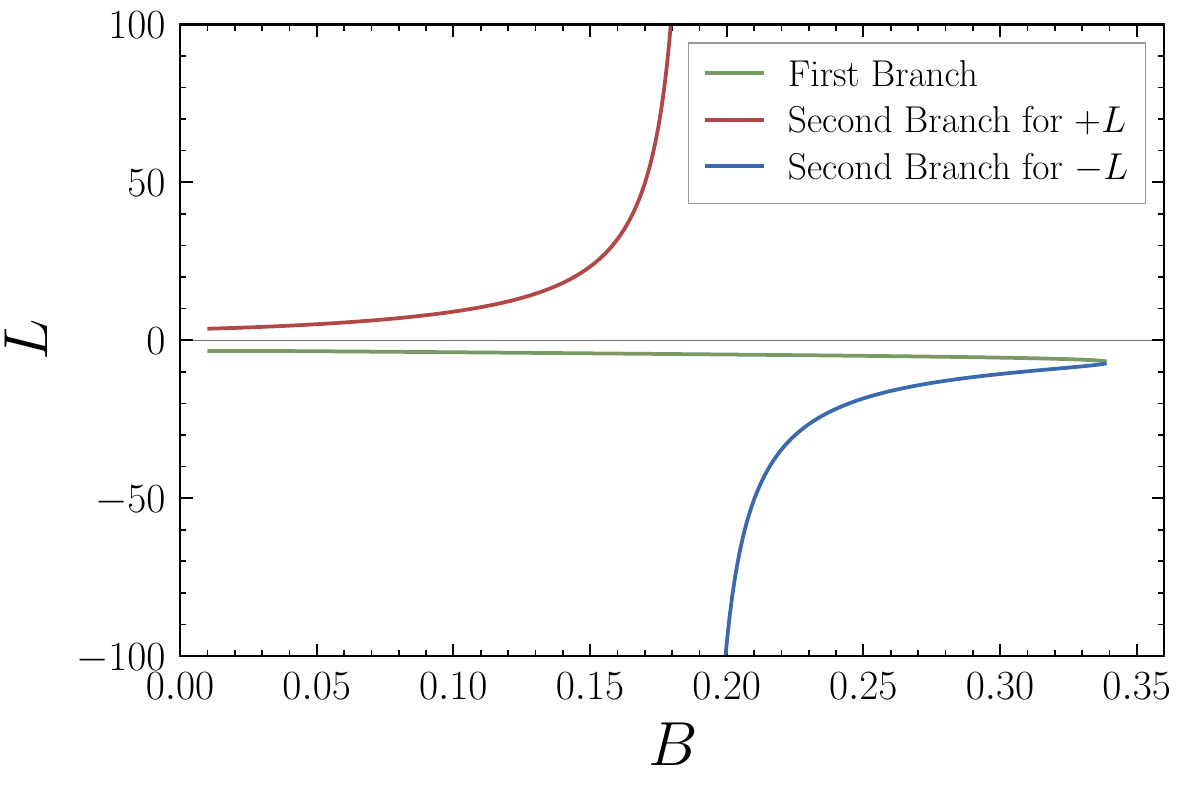}
  \caption{{The variation of $L$ with respect to $B$ from $0$ to $0.34$ with a fixed $e=-2$ for ISCOs. }}\label{BL}
\end{figure}

To make sure the existence of circular orbits, we should ensure the equation $\partial_rV=0$ has solutions outside the horizon $r_h=2$. Note that the right side of the Eq. (\ref{pvl}) is a quadratic polynomial for {$L$}, thus, we have the discriminant of the quadratic equation
\bea
\Delta=16Y^2 e^2 B^2 r^4-16XZr^2(4+B^2r^2)^2\ge0\,,
\eea
holds {for at least one radius $r>2$}, which gives us a constraint on $e$ and $B$. This inequality is too complicated to solve analytically, so that, we numerically obtain the result shown as the red line in Fig. \ref{constraint}. {Under and including the red line, we have $\Delta\ge0$ for at least one radius $r>2$; above the red line, $\Delta<0$ for all $r>2$. If the quadratic coefficient vanishes, the degenerate equation must be checked separately.} In addition, we note that $Y\ge0$ and $Z\ge0$ always hold when $r>2$, as a result, when $L>0$ we can see that the latter two terms of the right side of the Eq. (\ref{pvl}) are positive. Therefore, if the $\partial_r V=0$ is satisfied, we conclude the first term of the right side of the Eq. (\ref{pvl}) must be negative which implies $X<0$ for $r>2$. From this condition, we find $B<B_c=\frac{2}{5}\sqrt{\frac{1}{15}\left(169-38\sqrt{19}\right)}\simeq0.189$ which is shown as the blue line in Fig. \ref{constraint}. Very interestingly, in \cite{Junior:2021dyw}, the authors found that $B_c$ is also a critical value of the magnetic field strength and no light rings \footnote{Roughly speaking, Light rings are circular photon orbits in a spacetime. We suggest readers to see \cite{Cunha:2017qtt, Cunha:2020azh, Guo:2020qwk, Guo:2021bcw, Wei:2020rbh} if interested in more properties of light rings in stationary spacetimes.} exists outside the apparent horizon for $B>B_c$ while both the stable and unstable light rings exist for $B<B_c$ . As mentioned above, we can come to the conclusion that when the values of $(e, B)$ are in the region between the red and blue line, we only have $L<0$ circular orbits. And in the region between $B=B_c$ and $B=0$, both $L>0$ and $L<0$ circular orbits exist. We give an example shown in Fig. \ref{BL} for $e=-2$.

\begin{figure}[!htbp]
  \centering
  \includegraphics[width=10.0cm]{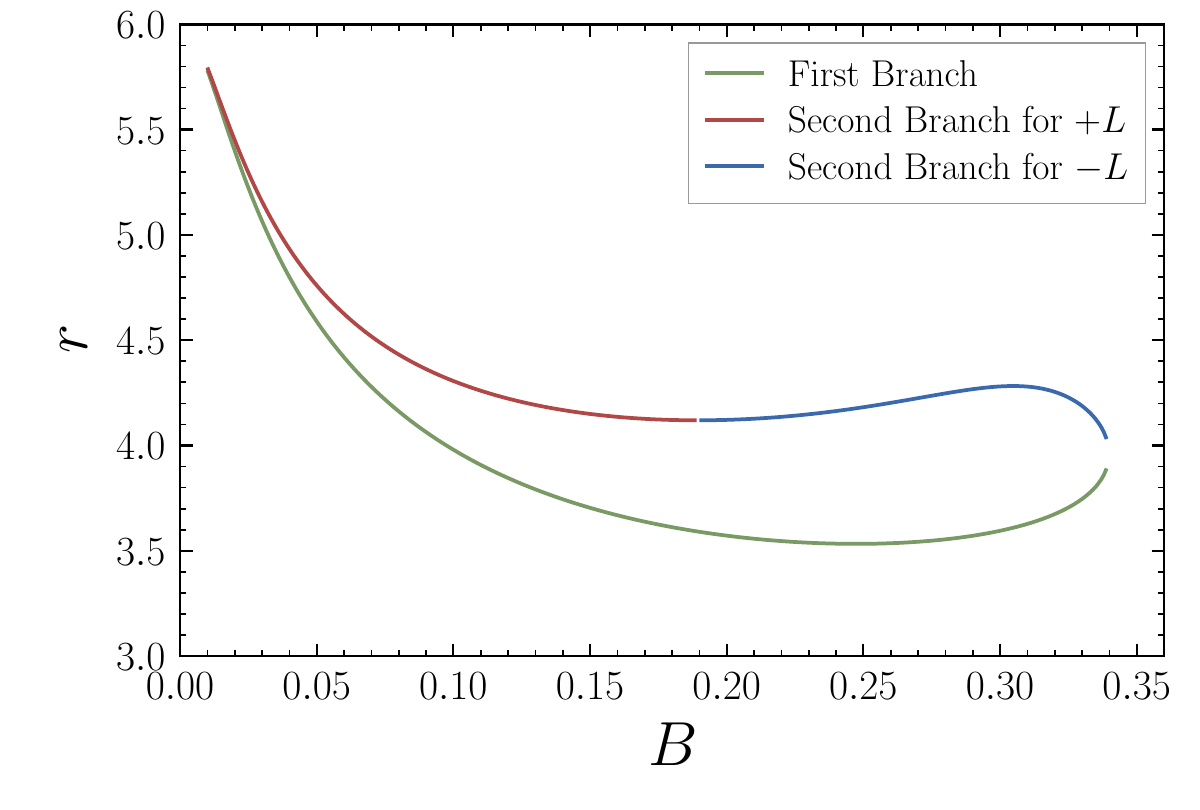}
  \caption{{The radius of the ISCO with respect to $B$ for $e=-2$. $B$ goes from $0$ to $0.34$.}}\label{BR}
\end{figure}

On the other hand, we can rewrite $\partial_rV$ in this form
\bea\label{pvr}
\frac{32r^4\partial_rV}{\Lambda}&=&3PB^4r^7-5PB^4r^6+4(8Q + 2 B^2 L^2 + 3P)B^2r^5-4(12Q + 3B^2 L^2 + 4P)B^2 r^4\nn\\
&&+16 B^2 L^2r^3+16(4Q+P)r^2-64L^2r+192L^2\,,
\eea
where
\bea
P=(-2 e + B L)^2\,,\quad\quad Q=1-e^2\,.
\eea
From $V=E^2$, we can obtain
\bea
\Lambda\left(\frac{L^2}{\Lambda^4\lambda^2f}-\frac{L^2}{r^2}+\frac{B^2L^2}{4\Lambda}\right)=\frac{P}{4}+\frac{Q}{\Lambda}\,,
\eea
where we have introduced a new parameter $\lambda^2=L^2/E^2$, thus, we can see that when $Q=0$, {after eliminating $P$ with $V=E^2$,} each term in the right hand of the Eq. (\ref{pvr}) contains a factor $L^2$ which can be dropped to solve $\partial_rV=0$. Therefore, we find that apart from the parameter $B$, $r$ is only related to $\lambda^2$, so that, it is interesting to find that in this case $Q=0$, that is $e=-1$ in our convention, no matter what values of $L$ and $E$ are, $|L/E|$ is the same for a circular orbit {at a fixed radius}. On the contrary, when $e\neq-1$, the value {$|L/E|$ is generally} different for a circular orbit.

{Within the parameter region allowed by Fig.~\ref{constraint}, we locate the ISCO by further imposing $\partial_r^2V=0$ and checking the radial stability of the neighboring circular orbits.} An example of the radius of ISCO with respect to $B$ is given in Fig. \ref{BR} with $e=-2$.
\section{Synchrotron radiations of charged particles in SMBH spacetime}\label{sec3}
In this section, we move to investigate the synchrotron radiations of charged particles orbiting the SMBH on the equatorial plane. The synchrotron radiations of charged particles have good polarization properties, and the polarization direction can reflect the feature of the magnetic fields surrounded the black hole. In order to find the polarized information of synchrotron radiations on the screen of observers, a possible way to know the polarizations at the position of the source.
In this work, we would like to focus on the study of polarization directions. {Following the magnetized-source model of~\cite{EventHorizonTelescope:2021btj,Gelles:2021kti}, we assume that the emitted polarization is perpendicular to both the photon momentum and the magnetic field in the emitter's comoving frame. The corresponding polarization vector is given by}
\bea\label{ffeq}
f^{\mu}=N^{-1}\epsilon^{\mu\nu\rho\sigma}Z_\nu k_\rho b_\sigma\,,\qquad
b^\mu={}^\ast F^{\mu\nu}Z_\nu\,,
\eea
where $N$ is the normalized factor which is not important to the direction, $k_\alpha$ is the $4$-momentum of the radiations, $Z^\alpha$ is the $4$-velocity of the charged particles, and $b^\mu$ is the magnetic field measured by the emitter. We choose $N$ such that $f\cdot f=1$. We consider only linear polarization, with Stokes $V=0$.

\subsection{Null geodesics in SMBH spacetime}

{In Sec. \ref{sec2}, we have obtained the $4$-velocity of charged particles in stable circular orbits on the equatorial plane.} The $4$-momentum $k^\alpha$ can be obtained considering null geodesics originating from the source at $(t_s, r_s, \theta_s=\pi/2, \phi_s)$ and reaching the observer at $(t_o, r_o, \theta_o, \phi_o)$. Note that the equation of motion along null geodesics cannot be separated in SMBH spacetime, one has to employ the numerical ray-tracing method to find the trajectory connecting the source and the observer and determine $k^\alpha$ at the source.

\begin{figure}[!htbp]
  \centering
  \includegraphics[width=8.0cm]{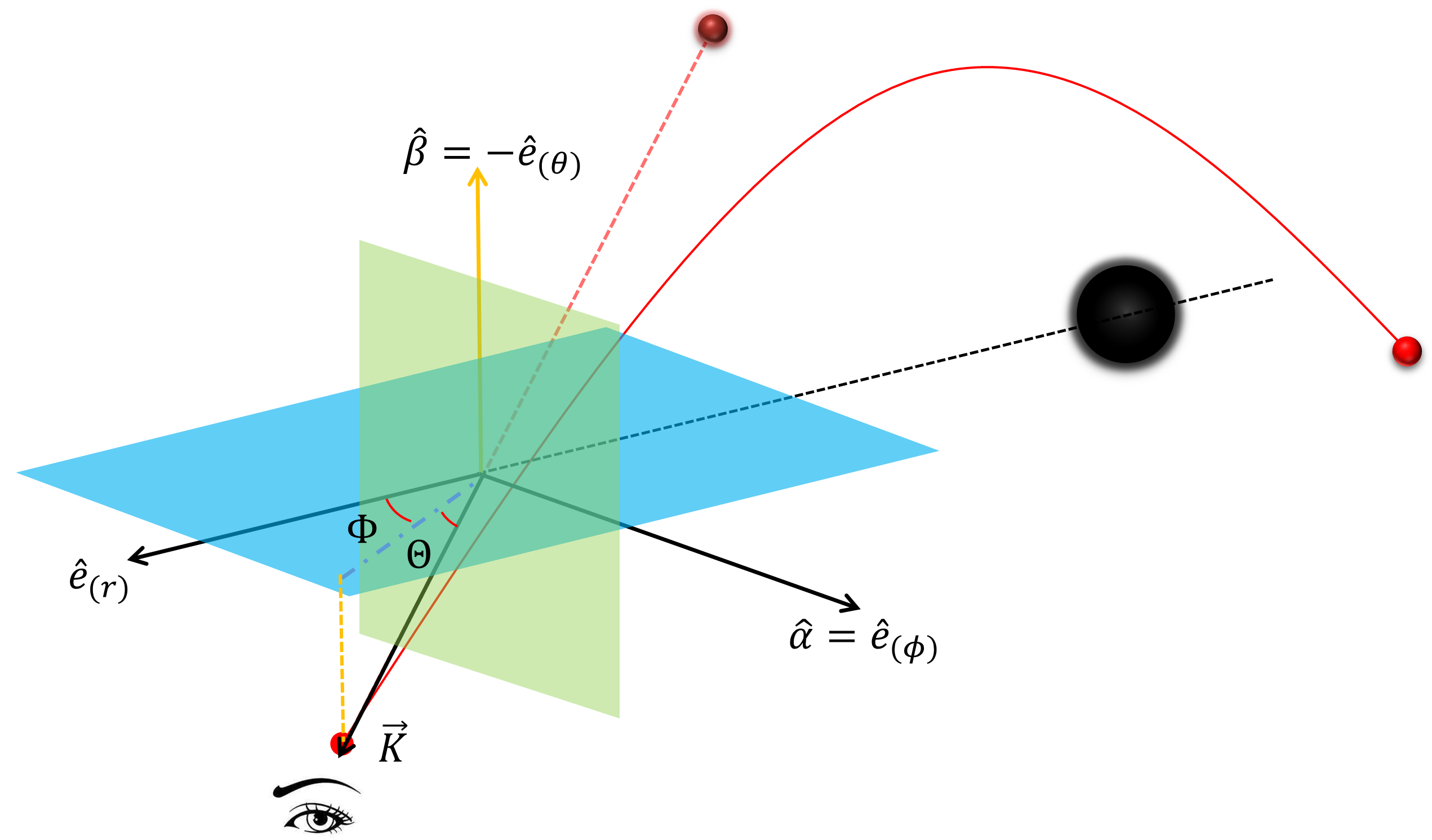}
  \caption{The main content of the picture is taken from \cite{Li:2020drn}, which shows the projection of the photon’s momentum $\vec{K}$ in the observer’s frame. $(\hat{x}, \hat{y})$ are the Cartesian coordinates glued to the screen of the observer. }\label{graph}
\end{figure}

Similar with the timelike particles, the components of $4$-momentum $k^\mu$ in the {Schwarz\-schild-like coordinates} takes
\bea
k^\mu=(t^\prime, r^\prime, \theta^\prime, \phi^\prime)\,,
\eea
where $\prime$ denotes derivative with respect to {an affine parameter} $s$. In SMBH spacetime, in order to do the numerical geodesic evolution more conveniently, we obtain the null geodesic equations in the Hamiltonian canonical formulism
\bea\label{geh}
k^\prime_\mu=-\frac{\partial H}{\partial x^\mu}\,,\quad\quad x^{\mu\prime}=\frac{\partial H}{\partial k_\mu}
\eea
where $H$ and $x^\mu$ are the Hamiltonian and positions of the photons, respectively. {We use $H=\tfrac12 g^{\mu\nu}k_\mu k_\nu=0$.} In particular, considering the Killing vectors $\partial_t$ and $\partial_\phi$ in SMBH spacetime, we are allowed to define the constants along the null geodesics
\bea
\omega=-k_t=\Lambda^2ft^\prime\,,\quad\quad l=k_\phi=\frac{r^2\sin^2\theta}{\Lambda^2}\phi^\prime
\eea
which are the energy and angular momentum, respectively.

Since optical paths are reversible, we can solve the Eq. (\ref{geh}) using backward ray-tracing method, that is, we can take $(x^\mu, -k_\mu)$ observed at the observer as the initial values of the Eq. (\ref{geh}). In order to achieve this aim, at first we build a local rest frame in the neighborhood of the observer, of which the tetrad reads
\bea\label{observer_tetrad}
\hat{e}_{(t)}=\Lambda^{-1}f^{-1/2}\partial_t\,,\hat{e}_{(r)}=\Lambda^{-1}f^{1/2}\partial_r\,,\hat{e}_{(\theta)}=\Lambda^{-1}r^{-1}\partial_\theta\,,\hat{e}_{(\phi)}={\Lambda}r^{-1}\sin^{-1}\theta\partial_\phi
\eea
where $\hat{e}_{(t)}$ is timelike and the others are spacelike. The $4$-momentum $k^\mu$ can be measured by the observer and we find
\bea\label{ckh}
k^{(t)}=-k_\mu \hat{e}^\mu_{(t)}\,,\quad\quad k^{(i)}=k_\mu \hat{e}^\mu_{(i)}\,,i=r, \theta, \phi.
\eea
In addition, we introduce the $3$-momentum vector $\vec{K}$ of $k^\mu$, seen as in Fig. \ref{graph}. Since $k^\mu$ is a null vector, we have $|\vec{K}|=k^{(t)}$. Then on the screen of the observer, we can define the apparent position of the image of the photons in the Cartesian coordinates $(\alpha, \beta)$ as
\bea\label{albe}
\alpha\equiv-r_o\tan\Phi\,,\quad\quad {\beta\equiv r_o\frac{\tan\Theta}{\cos\Phi}}.
\eea
where, the angles $\Theta$ and $\Phi$ are defined as
\bea\label{thph}
\sin\Theta=\frac{{k^{(\theta)}}}{|\vec{K}|}\,,\quad\quad\tan\Phi=\frac{k^{(\phi)}}{k^{(r)}}\,,
\eea
Therefore, plunging the Eq. (\ref{thph}) into the Eq. (\ref{albe}), we can obtain the final expressions of $\alpha$ and $\beta$ in this form
\bea\label{screen_revised}
\alpha=-r_o\frac{k^{(\phi)}}{k^{(r)}}\,,\quad\quad{\beta=r_o\frac{k^{(\theta)}}{k^{(r)}}}\,.
\eea
Then, inversely combining with the Eq. (\ref{ckh}) we can find the components of the $4$-momentum $k^\mu$,
\bea\label{camera_ray}
k^t&=&\Lambda^{-2}f^{-1}\omega\,,\nn\\
k^r&=&{\frac{r_o}{D}\Lambda^{-2}\omega}\,,\nn\\
k^\theta&=&{\frac{\beta}{r_oD}\Lambda^{-2}f^{-1/2}\omega}\,,\nn\\
k^\phi&=&{-\frac{\alpha}{r_oD\sin\theta_o}f^{-1/2}\omega}\,,
\eea
in the coordinates $(t, r, \theta, \phi)$. {Here $D=\sqrt{r_o^2+\alpha^2+\beta^2}$.} Now, we have obtained all the initial values $(x^\mu, -k_\mu)$ at the observer. Then we are able to solve the null geodesics using the Hamiltonian canonical formulism numerically and find the momentum vector $k^\mu$ at the source $(t_s, r_s, \theta_s, \phi_s)$. It needs to be emphasized that the light rays may cross the equatorial plane many times before reaching the source. We introduce $m$ to denote {the number of equatorial crossings before the selected emitting crossing} and we can see that the value of $m$ will give the $m+1$th image on the observer's screen. In the present work, we would like to show the primary and secondary images of the source.

\subsection{Polarization propagation {by parallel transport} in SMBH spacetime}
In this subsection, we move to calculate {the observed polarization by parallel transport} along null geodesics in SMBH spacetime.
\par\begingroup
The Schwarzschild--Melvin spacetime is generically of Petrov type I~\cite{Ortaggio:2004melvin}. We calculate the observed polarization by numerical parallel transport~\cite{10x}, neglecting absorption and Faraday effects. At the observer, we construct two orthonormal screen vectors $s_A^\mu$, $A=\alpha,\beta$, perpendicular to the observed photon direction and transport them backwards along $\ell^\mu=dx^\mu/d\bar s=-k^\mu$:
\bea\label{parallel_transport}
\frac{ds_A^\mu}{d\bar s}=-\Gamma^\mu{}_{\nu\rho}\ell^\nu s_A^\rho\,,\qquad
\ell\cdot s_A=0\,,\qquad s_A\cdot s_B=\delta_{AB}\,.
\eea
At the source, we project the polarization vector in Eq.~\ref{ffeq} onto the transported screen basis. This gives the polarization components and angle on the observer's screen:
\bea\label{evpa_revised}
F_A=f_\mu s_A^\mu\,,\qquad
\chi=\operatorname{atan2}(F_\beta,F_\alpha)\pmod\pi\,.
\eea
\par\endgroup
\section{Results}\label{sec4}
In this section, we are ready to find out the results of the polarized images of the charged source orbiting the SMBH on the equatorial plane. In our work, on one hand, we would like to consider $B$ is small so that the radius of the ISCO in SMBH spacetime is close to the radius of the ISCO in Schwarzschild black hole spacetime. On the other hand, as found in \cite{Junior:2021dyw}, the shadow and image become weird when $B>B_c$, thus, we take $B=0.02$ in the following.

\begin{figure}[H]
\centering
\includegraphics[width=.96\linewidth]{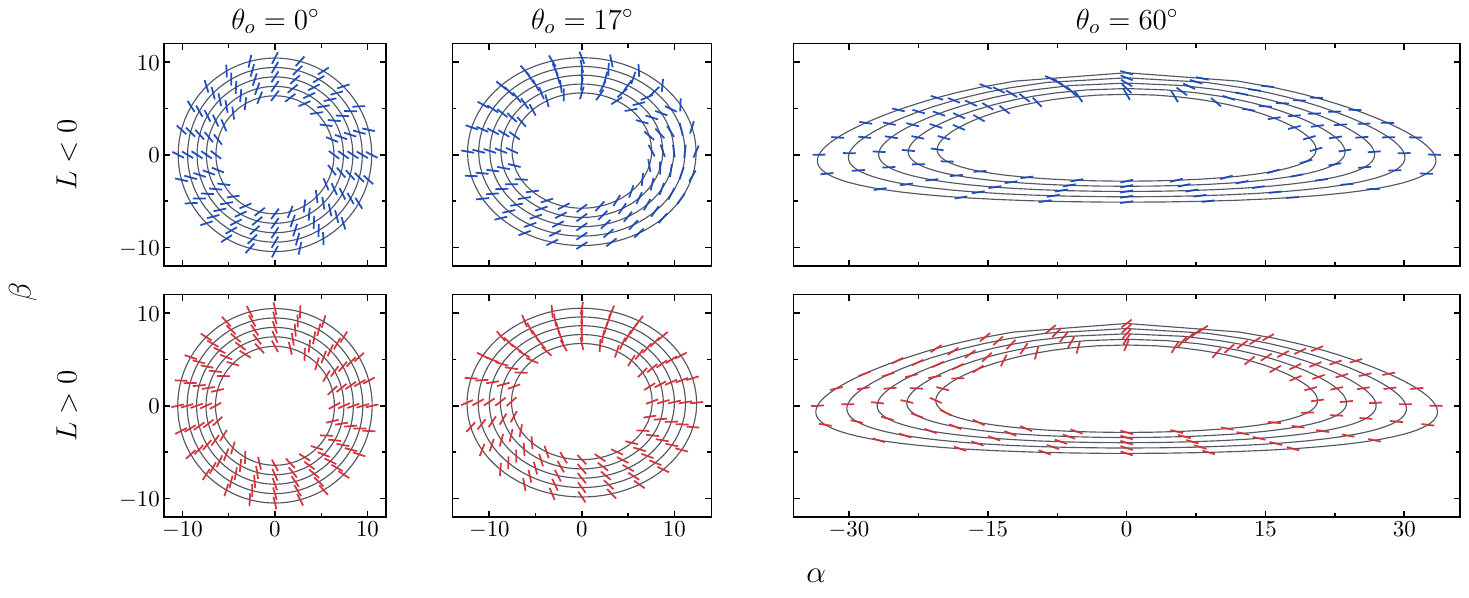}
\caption{{}Primary polarized images of charged particles orbiting around the SMBH on the equatorial plane observed at the inclination angle $\theta_o=$0\degree, 17\degree{} {and 60\degree}. We set $e=-2$ {and $r_o=100M$}. The radii of ISCO are {$5.40$} for $L<0$ and {$5.45$} for $L>0$, which are shown as the innermost circles in the plot, respectively. The outer circle next to the innermost one is $\rI+1$ and so on. Five orbits are given for each case in the plot. {The upper and lower rows correspond to $L<0$ and $L>0$, respectively.}}\label{e2theta017}
\end{figure}

In Fig. \ref{e2theta017}, we give the primary polarized images of orbiting charged particles with the charge being $e=-2$ on the equatorial plane in the SMBH spacetime observed by observers at the inclination angle $\theta_o=${0\degree, 17\degree{} and 60\degree}. The radius of ISCO is {$5.40$} for charged particles with a negative $L$ while for a positive $L$, the ISCO is located at {$\rI=5.45$} which is a little bigger than that for a negative $L$. From these polarized images, we can draw some interesting conclusions. At first, we can see that the image of charged particle becomes oblate when the inclination angle $\theta_o$ increases. Similar observations of black hole shadows can be also found in \cite{Hu:2020usx}, where the non-rotating black holes immersed in Wald magnetic fields are considered. Secondly, we find that the polarization directions are different for $L<0$ and $L>0$ orbits at the same inclination angle. This finding can help us to distinguish the {sign} of the angular momentum of equatorial orbits of charged particles through their polarization directions of synchrotron radiations. Moreover, one can see that {the polarization directions change as the inclination angle $\theta_o$ increases}. In addition, our results reveal {differences} of the polarization directions between the inner orbit and the outer orbit {at comparable positions along their image tracks}.

\begin{figure}[H]
\centering
\includegraphics[width=.96\linewidth]{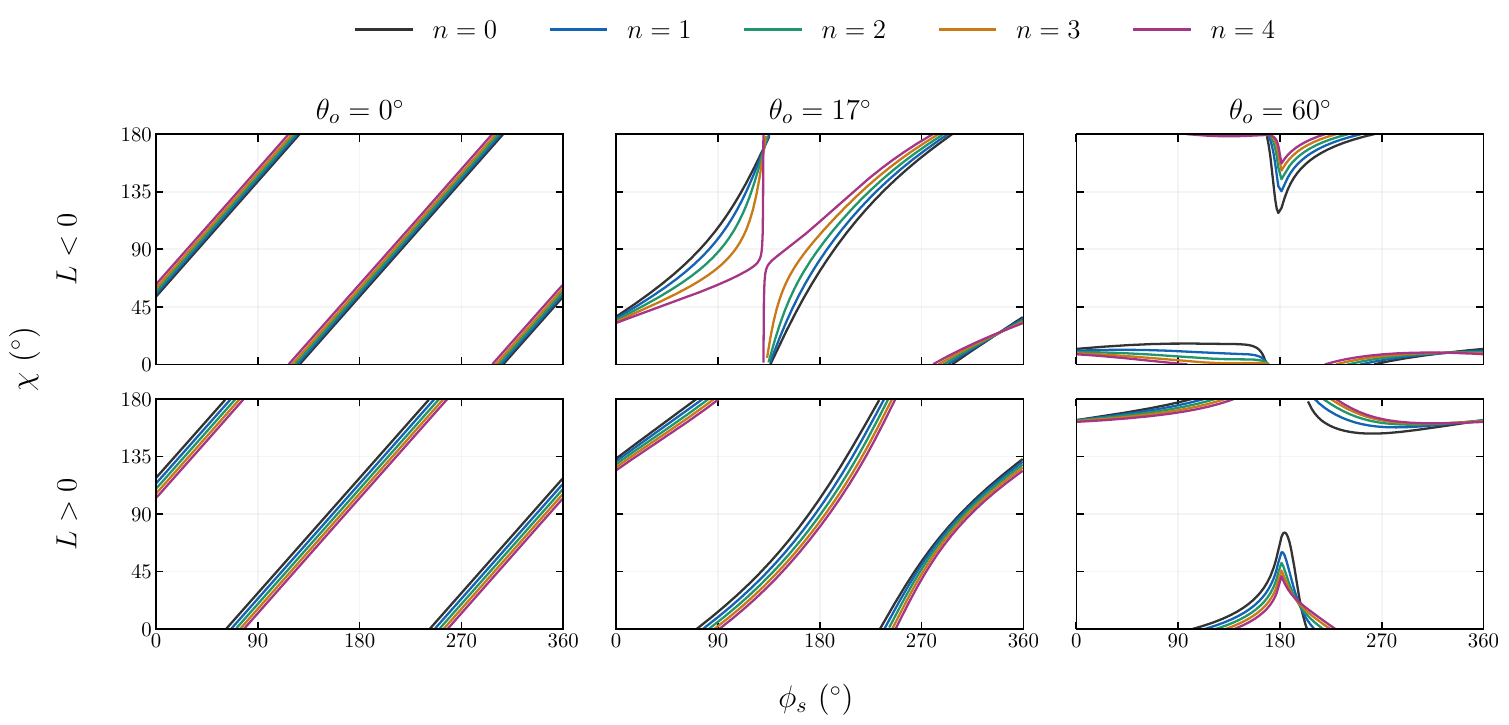}
\caption{{Observed polarization angle $\chi$ as a function of source azimuth $\phi_s$ for the primary images in Fig.~\ref{e2theta017}. Colors denote the source radii $r=\rI+n$, with $n=0,1,2,3,4$. The panel arrangement is the same as in Fig.~\ref{e2theta017}. The angle is shown modulo $180^\circ$; breaks at this boundary do not indicate physical jumps.}}\label{primary_evpa}
\end{figure}

In Figure~\ref{primary_evpa}, we show the dependence of the observed polarization angle on source azimuth. At $\theta_o=60$\degree{}, the curves intersect near $\phi_s=170$\degree{} for $L<0$ and $\phi_s=195$\degree{} for $L>0$, showing that the ability of polarization to distinguish source radii depends on source azimuth.

\begin{figure}[H]
\centering
\includegraphics[width=.72\linewidth]{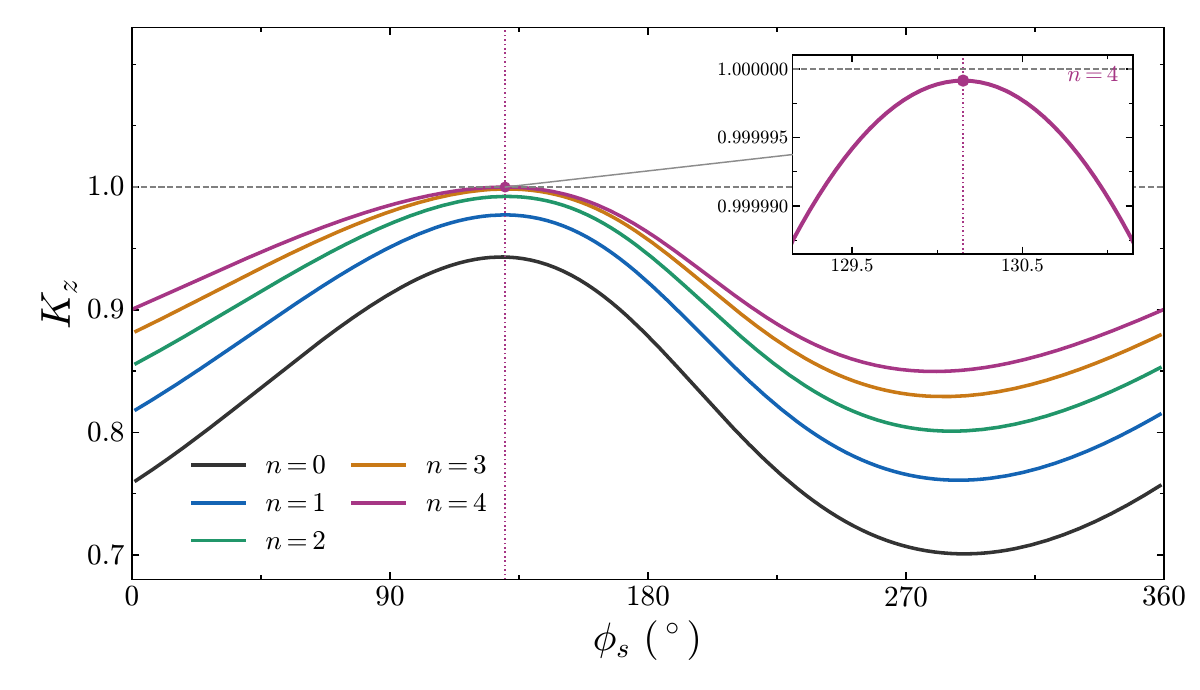}
\caption{{Normalized photon momentum component $K_z$ along the magnetic field in the emitter's comoving frame for the $L<0$ primary images at $\theta_o=17$\degree. The source radii are $r_s=\rI+n$; colors and other parameters are as in Fig.~\ref{primary_evpa}. The inset shows the peak for $n=4$.}}\label{photon_alignment}
\end{figure}

A rapid but continuous variation of the EVPA occurs near $\phi_s=130$\degree{} for the outermost orbit ($n=4$) in the upper middle panel of Fig.~\ref{primary_evpa} ($\theta_o=17$\degree{}, $L<0$). To understand this behavior, in Fig.~\ref{photon_alignment}, we show the normalized photon momentum component $K_z=k^{\hat z}/|\boldsymbol{k}|=\cos\zeta$, measured in the emitter's comoving orthonormal frame with the $z$-axis along the magnetic field. For the outermost orbit, $K_z$ reaches its maximum near $\phi_s=130.15$\degree, where the photon momentum is closest to parallel to the magnetic field, with a minimum angle $\zeta\simeq0.075$\degree. The transverse momentum component is then very small. Only this component contributes to the cross product with the magnetic field, so small changes in its components can produce a large rotation of the emitted polarization axis. This rapid variation is carried to the observer by parallel transport and appears in the observed EVPA.

\begin{figure}[H]
\centering
\includegraphics[width=.96\linewidth]{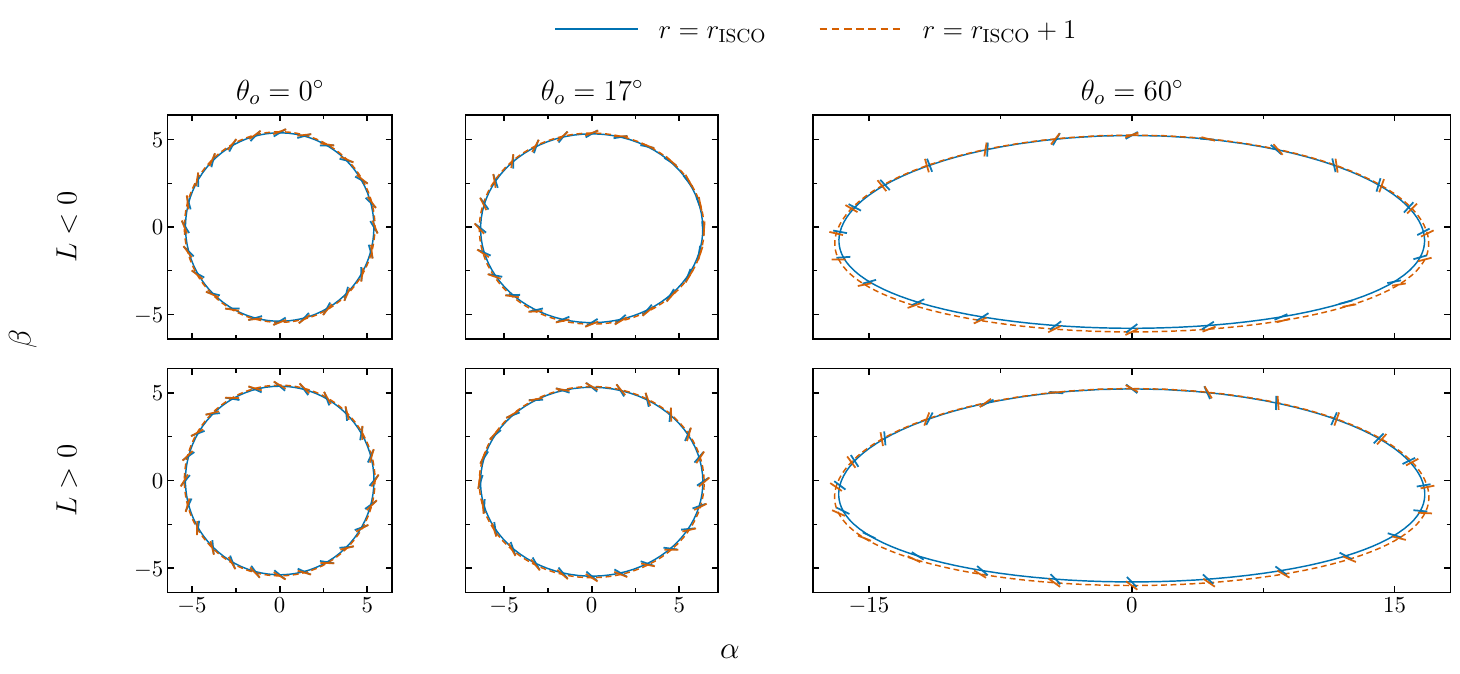}
\caption{{}Secondary polarized images of charged particles orbiting around the SMBH on the equatorial plane observed at the inclination angle $\theta_o=$0\degree, 17\degree{} {and 60\degree}. The other parameters are the same with those  in Fig. {\ref{e2theta017}}. {Blue solid and orange dashed curves show the $\rI$ and $\rI+1$ orbits, respectively. The upper and lower rows correspond to $L<0$ and $L>0$, respectively.}}\label{se2theta123}
\end{figure}

In Fig. \ref{se2theta123}, we give the secondary polarized images of orbiting charged particles with the charge being $e=-2$ on the equatorial plane in the SMBH spacetime observed by observers at the inclination angle $\theta_o=${0\degree, 17\degree{} and 60\degree}. The difference of the radii of two adjacent orbits is $1$ and we {can} see that the images of two adjacent orbits are very close on the $\alpha-\beta$ plane, some parts even overlap when the inclination angle $\theta_o=$60\degree, thus, we would like to show only two orbits for each plot so that it would help people to better tell apart the images and corresponding polarization directions.

\begingroup\interlinepenalty=10000
From these plots, we can find the secondary images and corresponding polarization directions of charged particles behave similarly with the primary images.\par\endgroup

\begin{figure}[H]
\centering
\includegraphics[width=.96\linewidth]{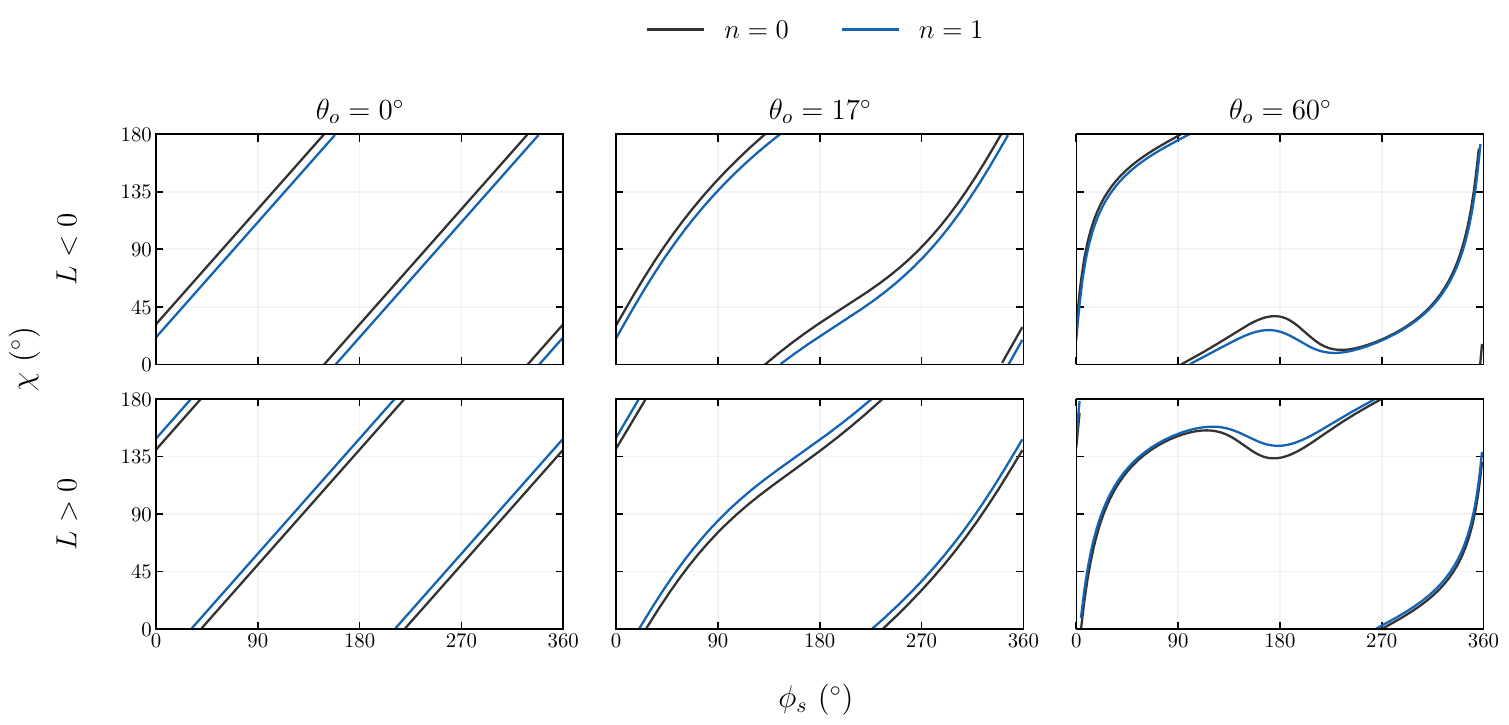}
\caption{{Observed polarization angles for the secondary images in Fig.~\ref{se2theta123}, with $r=\rI+n$ and $n=0,1$. The panel arrangement is the same as in Fig.~\ref{se2theta123}, and the angle convention is the same as in Fig.~\ref{primary_evpa}.}}\label{secondary_evpa}
\end{figure}

{\clubpenalty=10000\widowpenalty=10000
{At $\theta_o=17$\degree, the secondary images of adjacent source orbits are much closer together than their primary images. Figure~\ref{secondary_evpa} shows differences in their polarization angles that vary with source azimuth. These differences can help distinguish the source orbits even when their secondary image tracks are nearly coincident.}\par
}

\begin{figure}[H]
\centering
\includegraphics[width=.72\linewidth,height=.46\textheight,keepaspectratio]{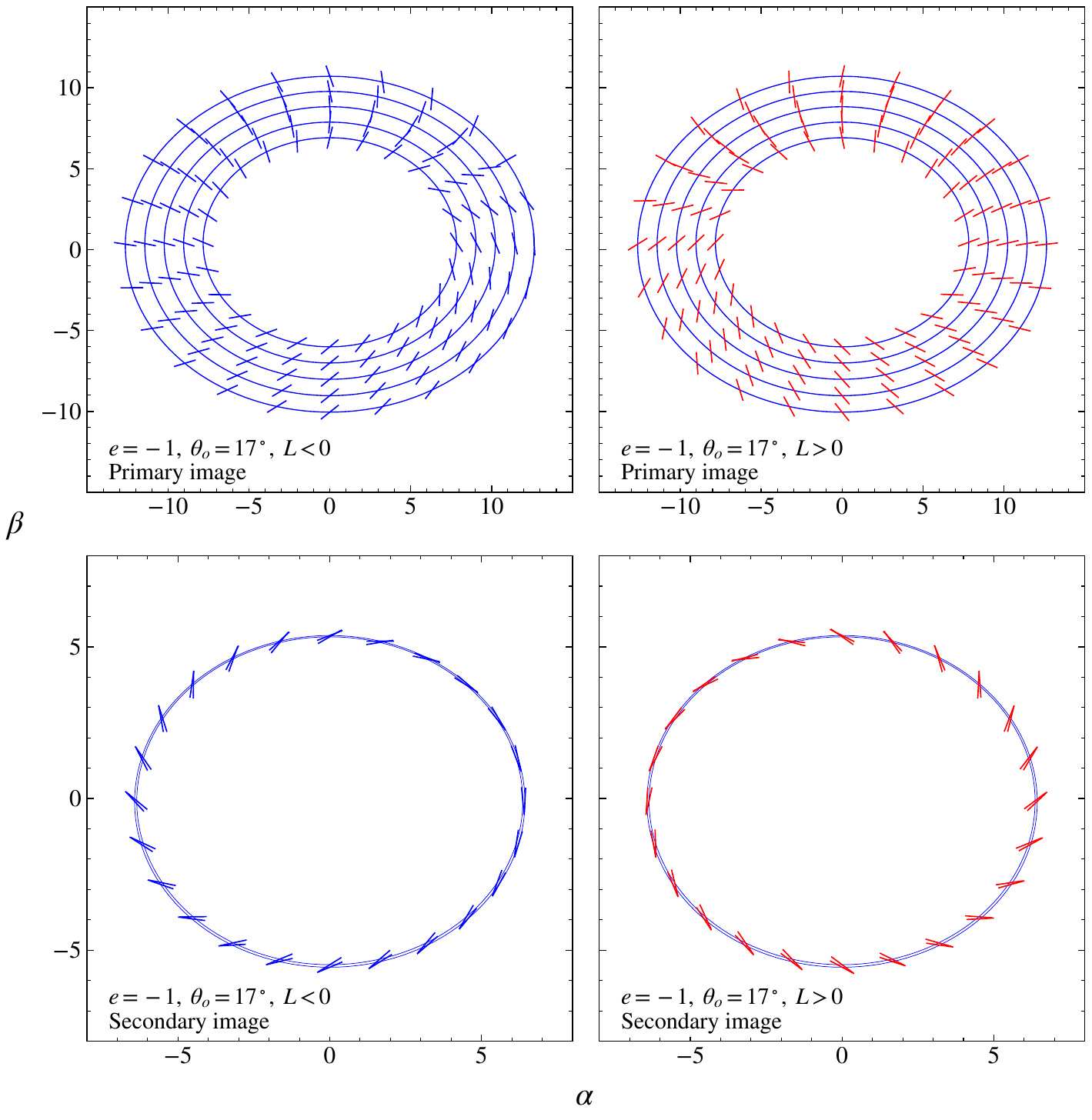}
\caption{{}Primary and secondary polarized images of charged particles orbiting around the SMBH on the equatorial plane observed at the inclination angle $\theta_o=$17\degree. The charge of the particle is $e=-1$ and the radius of the ISCO is {$5.63820$}. We show five orbits for primary images and two orbits for secondary images. The difference of radii of two adjacent orbits is $1$.}\label{e1theta17}
\end{figure}

Recall that the circular orbits have a same $|L/E|$ {at the same radius} when the strength of the magnetic field is given and the charge of the particle takes $e=-1$. We would like to see the polarized images of orbiting charged particles. In Fig. \ref{e1theta17}, we show the primary and secondary polarized images of charged particles in SMBH spacetime at $\theta_o=$17\degree. The radius of the ISCO in SMBH spacetime is {$5.63820$}.

{\interlinepenalty=10000
{As shown in Fig.~\ref{e1theta17}, the $L>0$ and $L<0$ branches at the same orbital radius produce coincident image tracks but distinct polarization patterns. Their different orbital velocities lead to different photon propagation directions in the emitter's comoving frame and hence different emitted polarization directions. This allows the two branches to be distinguished through polarization even when their image tracks coincide.}\par
}
\section{Summary and Discussion}\label{sec5}
In this paper, we have revisited the ISCO of charged particles and studied the polarized images of synchrotron radiations of charged hotspots orbiting on the equatorial plane in SMBH spacetime based on {a model in which the emitted linear polarization is linked to the local magnetic-field geometry}. For the part of ISCOs of charged particles, we found the {constraint} of the strength of the magnetic field $B$ {for circular-orbit existence}. Interestingly, we identified a critical $B_c$ that only $L<0$ orbits survive {over} $B_c$ and there are both $L>0$ {and} $L<0$ orbits {under} $B_c$ {in the allowed circular-orbit domain}. In particular, our critical $B_c$ is exactly the critical value appeared in \cite{Junior:2021dyw}, where they found there are no light ring when $B>B_c$ and light rings exist when $B<B_c$.

For the part of synchrotron radiations of charged hotspots, based on the formula in Eq. \ref{ffeq}, we obtained the polarized images of equatorial orbiting charged particles employing the numerical backward ray-tracing method and {numerical parallel transport} in SMBH spacetime. Some general characteristics of the shape and the polarizations of images were summarized in the last section. {We further examined the observed polarization angle as a function of source azimuth. Polarization helps distinguish adjacent source orbits with nearly coincident secondary image tracks at $\theta_o=17$\degree{}, as well as the two $e=-1$ orbital branches whose image tracks coincide. For nearly aligned photon and magnetic-field directions in the emitter's frame, we find a rapid but continuous variation of the observed polarization angle with source azimuth. This behavior arises from the rapid rotation of the small transverse photon-momentum component that determines the emitted polarization axis.} It's worth noting that our results cannot be directly comparable with the results in \cite{EventHorizonTelescope:2021btj}, where they considered the polarized images of fluids on the equatorial plane and the magnetic field configurations can be arbitrarily given. In particular, their fluids can be assumed to include charged particles in a complex state of motion, {rather than the charged circular orbits considered here}. However, although our model is simple, all the {assumed} physical mechanisms are clear and can play a theoretical base of understanding synchrotron radiations near black holes.

\section*{Acknowledgments}
We are grateful to Yehui Hou and Zezhou Hu for helpful discussions. We thank Haopeng Yan for reading the manuscript of this paper and for useful comments. The work is in part supported by NSFC Grant  No. 12205013 and 12575048. M.G. also acknowledges support from the BNU Tang Scholar Program.
\bibliographystyle{utphys}
\bibliography{note}

\end{document}